# A gaseous stripper proposal based on hadrontherapy facility HIMM

XIE Xiu-Cui(谢修璀)[1,2]　　SONG Ming-Tao(宋明涛)[1]　　ZHANG Xiao-Hu(张小虎)[1,2]

1 Institute of Modern Physics, Chinese Academy of Sciences, Lanzhou 730000, China

2 University of Chinese Academy of Sciences, Beijing 100049, China

**Abstract:** Multi-turn Injection scheme with gaseous stripper is usually used in high intensity and super heavy ion injection process. With its advantage of long lifetime and uniformity, a gaseous stripper is proposed based on the under construction hadrontherapy facility HIMM (Heavy Ion Medical Machine). In this paper, the physical process between the injecting beam and the gaseous target is studied, and a simulation work is conducted based on the former developed code.

**Key words:** HIMM, injection, gaseous stripper, carbon ion, nitrogen target

**PACS:** 29.27.Ac, 29.20.Lq   **DOI:**

## 1 Introduction

Stripping process with gaseous stripper is an key arrangement in high intensity and super heavy ion injection because gaseous stripper is free from lifetime problem while a solid foil stripper may easily broken under the heat which is brought by the intense bombardment of the injecting beam.[1] A recent research shows that a carbon foil of 300μg/cm² will broke in 12 hours under the 1 pμA uranium beam.[2] Also, a gaseous stripper is much more easy to achieve uniformity because the typical foil thickness used in stripping is several tens of μg/cm² to several mg/cm², which is equivalent to a gaseous length of several cm to several tens cm. With all these advantage, as well as its significant potential usage in future large hadron accelerator facility motivate us to develop a gaseous stripper based on the currently under construction hadrontherapy facility HIMM.

## 2 physical process

The physical principle that happened between the injecting beam and the gaseous target is dominated by the dynamic balance between electron capture and loss process, which decide the charge state distribution.[1] The relevant impact include energy loss, angular scattering and position change, which is quite the same as in a solid foil stripper. To describe the process, many theoretical calculation and experimental result has been presented. Due to the extreme complexity of the collision process, a theoretical calculation based on simplified model shows difficulty in describing the interaction relation, so a phenomenological semiempirical method will be adopted.[3]

### 2.1 capture and loss cross section

When the injecting ions colliding with the target gas, it will capture/loss one electron or several electrons simultaneously at each collision. The cross section provide a fundamental basis to describe ion-target encounters:[4]

$$\sigma_L = 4\pi a_0^2 Z^{-2}(Z_T^2 + Z_T)(v_0/v)^2 \quad (1)$$

$$\sigma_C = 4\pi a_0^2 Z^5 Z_T^{1/3}(v_0/v)^6 \quad (2)$$

Where σ is cross section, the subscript L and C means loss and capture cross section respectively. Z and $Z_T$ are the atomic number of projectile particle and target particle, $v_0$ and $a_0$ represent Bohr velocity (2.188×10⁶m/s) and Bohr radius (5.291×10⁻⁷m), v denotes the velocity of the projectile particle.

### 2.2 charge state distribution

Received 29 October 2013

\* Supported by State Key Development Program of (for) Basic Research of China (2014CB845500)

The average equivalent charge state can be derive from the capture and loss cross section:

$$\sum_{\substack{q' \\ q' \neq q}} [F_{q'}\sigma(q',q) - F_q\sigma(q,q')] = 0 \quad (3)$$

Where $F_q$ is the fraction of particle in q state and σ(q,q') denotes the cross section that from charge state q to q'.

Also, the width of the charge distribution could be defined by:

$$d = [\sum_{q'}(q' - \overline{q})^2 F(q')]^{1/2} \quad (4)$$

Where $\overline{q} = \sum_q qF(q)$.

With the help of the above equation, we introduce a semiempirical description by Dmitriev and Nikolaev:[4]

$$\frac{i}{Z} = \frac{\lg(v/z^{\alpha_1}/m)}{\lg(nz^{\alpha_2})} \quad (5)$$

$$d = d_0 Z^k \quad (6)$$

Where α₁, α₂, m, n and k are semiempirical constant based on particular type of target gas. Then we have equilibrium charge state relation showing in Fig. 1:

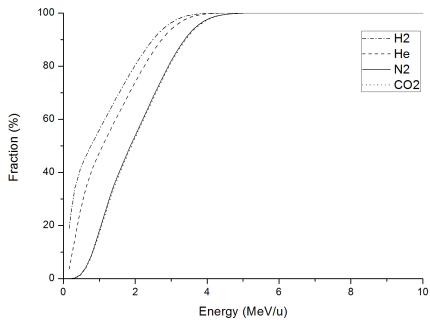

**Fig.1 the equilibrium charge state of $H_2$, He, $N_2$ and $CO_2$.**

Where we calculate four different target gas ($H_2$, He, $N_2$ and $CO_2$) under 7MeV/u carbon beam of charge state 5+.

**2.3 equivalent target thickness**

The thickness of the target gas has to be thick enough to allow the projectile beam to achieve equivalent distribution. We have:

$$x = NLP/(RT) \quad (7)$$

Where N is the Avogadro's number, L is the length of the cell, P and T are the pressure and temperature of the target gas, respectively, and the R is gas constant.

The unit that measure the target thickness x is molecules per $cm^2$, and the density effect need to be considered, too. According to the collision theory, the electron of the projectile ions would be stripped away or be excited to a higher potential state. If the density of the gas is thin enough, the excited electron may go back to the ground state before the next collision, while the thick gas may allow the excited electron to be stripped away, which means less energy is need, and this effect certainly change the thickness that needed by injecting beam to reach equivalent charge state. So we use the LISE code to calculate the equivalent thickness as showing in Fig.2.[5]

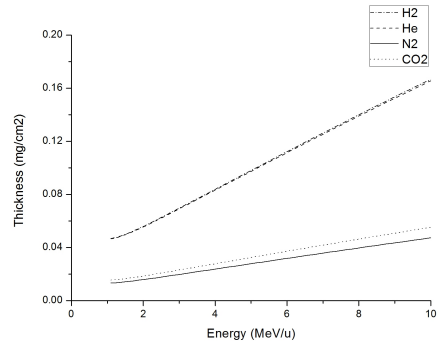

**Fig.2 the equilibrium target thickness of $H_2$, He, $N_2$ and $CO_2$.**

As can be seen from Fig.2, the thickness that needed by N2 is much smaller than lighter elements, along with the equilibrium charge state distribution that shows in Fig.1, we decide to choose N2 as the target gas.

**2.4 energy loss and angular scattering**

The energy loss of the injecting beam is described by the Bethe- Block formula, as in solid target. Fig.3 shows the average energy loss and its straggling calculated by LISE.

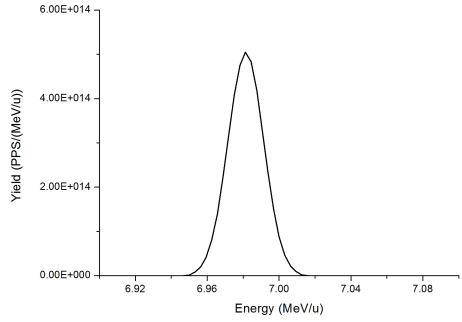

Fig.3 the energy loss distribution, the distribution type is Gaussian.

The angular scattering effect change the direction of the projectile beam, which could be calculated by LISE, too.

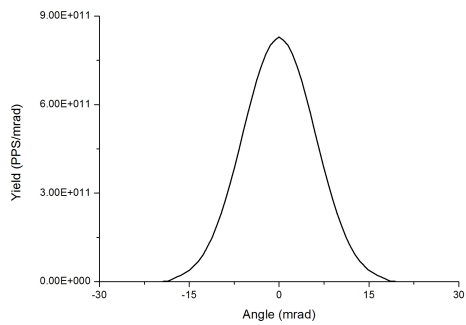

Fig.4 angular scattering distribution

With all these data obtained, we can simulate the injection process.

## 3 simulation

The simulation work is based on the former developed simulation code for charge exchange injection.[6] The code is modified to be used on the new multi-turn injection scheme.

### 3.1 schematic design and parameters

The nitrogen gas which used as gaseous stripper is kept in a differential pump system which located in the middle energy transport line between the cyclotron injector and synchrotron ring, as showing in Fig.5.[7]

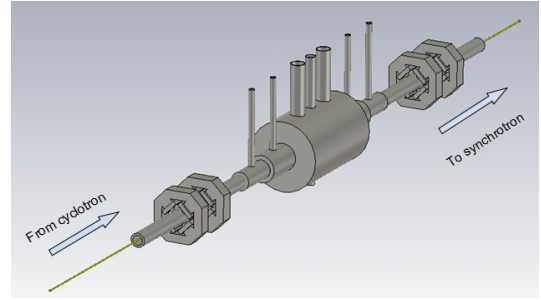

Fig.5 the schematic design of the gas stripper

Some of the basic parameters of the accelerator facility is summarized in Table 1.

Table 1 main parameters of the facility

| Injection particle | $C^{5+}$ |
|---|---|
| Injection energy | 7 MeV/u |
| Beam intensity | 10 μA |
| Synchrotron length | 56.173m |
| Ring intensity | $1\times10^9$ |
| Gas type | $N_2$ |
| Gas cell length | 100 mm |
| Gas cell pressure | 0.2 Atm |

### 3.2 simulation

The former developed simulation code needs to be modified to use on the new injection scheme. We input the new collision data as well as change the logical structure according to the multi-turn injection principle. The simulation process is showing in Fig.6.

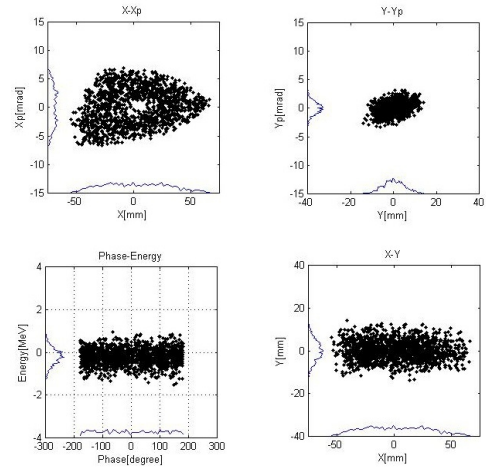

Fig.6 Multi-turn injection with gaseous stripper

The simulation results shows that only 15 times accumulation could be achieved, which is quite small compare to charge exchange injection.[7] Due to the low accumulation rate of the multi- turn injection scheme, we need a more powerful injector which can provide beam intensity at least 34.7μA. So, a linac injector made up of FRQ and DTL would become an alternate option.

**4 conclusion**

In this paper, a gaseous stripper proposal is designed and discussed based on the under constructing hadrontherapy facility HIMM. The interaction between the projectile beam and the target gas has been studied and appropriate semiempirical model has been chosen. A former developed charge exchange injection simulation code is modified to simulate the current multi- turn injection scheme. The results shows that 15 times accumulation could be achieved.